\documentclass[12pt,a4paper]{article}

\usepackage{newtxtext,newtxmath}

\usepackage[
a4paper,
left=2.5cm,
right=2.5cm,
top=2.5cm,
bottom=2.5cm
]{geometry}

\usepackage{setspace}
\usepackage{graphicx}
\usepackage{amsmath}
\usepackage{bm}
\usepackage{multirow}
\usepackage{makecell}

\usepackage{booktabs}
\usepackage{multirow}
\usepackage{tabularx}
\usepackage{array}
\usepackage[table]{xcolor}

\usepackage{caption}
\usepackage{subcaption}
\usepackage{float}
\usepackage{placeins}

\usepackage[linesnumbered,ruled,vlined]{algorithm2e}

\usepackage{enumitem}
\usepackage{microtype}
\usepackage{verbatim}

\usepackage[numbers,sort&compress]{natbib}

\usepackage[
colorlinks=true,
linkcolor=blue,
citecolor=blue,
urlcolor=blue
]{hyperref}

\usepackage{lineno}
\newcolumntype{C}{>{\centering\arraybackslash}X}
\newcolumntype{L}{>{\raggedright\arraybackslash}X}
\newcolumntype{R}{>{\raggedleft\arraybackslash}X}

\definecolor{tablerowcolor}{gray}{0.9}

\begin{document}

\title{\boldmath Improving Muon-Scattering Material Identification via Coarse Momentum Encoding and Unsupervised Domain Adaptation
}

\author{
Yuxin Bao$^{1,\dag}$,
Zhao Zhang$^{2,3,\dag}$,
Pei Yu$^{4,2}$,
Liangwen Chen$^{4,2,5,6,*}$,\\[0.4em]
Weibo He$^{7,*}$,
Yu Zhang$^{1,*}$,
Yuhong Yu$^{4,2,5,6}$,
Xueheng Zhang$^{4,2,5,6}$,\\[0.4em]
Lei Yang$^{4,2,5,6}$,
Zhiyu Sun$^{4,2,5,6}$
}

\date{}

\maketitle

\begin{center}
\small

$^1$ School of Physics, Hefei University of Technology, Hefei 230601, China

$^2$ Institute of Modern Physics, Chinese Academy of Sciences, Lanzhou 730000, China

$^3$ Frontiers Science Center for Rare Isotopes, Lanzhou University, Lanzhou 730000, China

$^4$ Advanced Energy Science and Technology Guangdong Laboratory, Huizhou 516000, China

$^5$ School of Nuclear Science and Technology, University of Chinese Academy of Sciences, Beijing 100049, China

$^6$ State Key Laboratory of Heavy Ion Science and Technology, Institute of Modern Physics, Chinese Academy of Sciences, Lanzhou 730000, China

$^7$ Institute of Materials, China Academy of Engineering Physics, Jiangyou 621907, China

\vspace{0.5em}

$^\dag$ These authors contributed equally to this work.

\vspace{0.5em}

* Corresponding authors: chenlw@impcas.ac.cn, njuyyf@163.com, dayu@hfut.edu.cn

\end{center}

\vspace{1em}

\begin{abstract}

Cosmic-ray muon scattering has shown considerable potential for detecting nuclear materials and other dense contraband, but practical deployment remains challenging. A major difficulty arises from the coupling between material properties and muon momentum, since the broad natural momentum distribution influences the scattering angle and prevents unambiguous material identification. In this work, we propose a Coarse Momentum–Aware Domain Adaptation (CMADA) method to enable precise identification of materials. Instead of relying on high-precision momentum measurements, the proposed framework adopts coarse momentum binning combined with unsupervised domain adaptation to learn transferable scattering representations. In addition, a precision review mode based on averaging repeated samplings was proposed to further enhances identification performance. The coarse momentum binning strategy improves same-domain identification accuracy from 62.15\% without momentum information to 89.52\% with 5-bin momentum information, and further to 93.37\% (precision review mode). Furthermore, the proposed unsupervised domain adaptation framework improves the cross-domain identification accuracy from 71.71\% for the source-only baseline to 89.00\% without requiring target domain labels. 

\end{abstract}

\noindent\textbf{Keywords:}
Unsupervised domain adaptation;
Muon scattering;
Material identification;
Momentum estimation.

\section{\label{sec:Introduction}Introduction}
Cosmic-ray muons originate from interactions between primary cosmic rays and atmospheric nuclei, producing extensive air showers of secondary particles \cite{neddermeyer1938cosmic}. Among these secondaries, muons are unusually penetrating due to their high energies and weak interactions. This exceptional penetration capability, which often exceeds that of conventional X-ray inspection, thereby enables non-destructive inspection of heavily shielded or large objects. This capability motivates Muon Scattering Tomography (MST), a novel and non-invasive imaging technique that estimates internal structure and material characteristics by measuring the multiple Coulomb scattering of muons traversing the object under inspection \cite{checchia2016review,wang2026u}. MST has been extensively studied across a range of fields, especially in nuclear security and customs screening, including detecting illicit nuclear materials in shielded cargo containers \cite{schultz2007statistical,jonkmans2013nuclear} and monitoring spent nuclear fuel for safeguards \cite{chatzidakis2016analysis}.

In recent years, significant progress has been made in leveraging cosmic-ray muon multiple Coulomb scattering for detecting and localizing dense, high-atomic-number (high-Z) objects \cite{hogan2004detection,vanini2019muography}. Subsequent studies further improved sensitivity by incorporating muon momentum information to better separate momentum-dependent scattering from material-dependent signatures \cite{bae2024nuclear,bae2022momentum,stapleton2014angle,yu2025improving}. Despite these advances, two key challenges remain for practical deployment. First, the broad momentum distribution of natural cosmic-ray muons introduces strong coupling between momentum-dependent scattering and material-dependent features, making accurate identification difficult without costly momentum measurement systems. Second, variations in measurement conditions across different inspection scenarios lead to domain shifts between training and testing data, significantly degrading model performance. Since it is impractical to collect labeled data for every scenario, unsupervised cross-domain transfer is essential.

However, most existing approaches remain limited to coarse material identification, typically distinguishing only low-, medium-, and high-Z categories. Such granularity is often insufficient for practical applications such as customs screening and nuclear security, where benign high-density cargo may exhibit scattering responses similar to those of controlled materials. In addition, low- and medium-Z materials dominate real inspection scenarios; finer identification among them can improve scene interpretation and reduce false alarms in mixed-material environments.

To address the domain shift issue, unsupervised domain adaptation offers a practical method by aligning feature distributions across domains without requiring target domain labels \cite{cicek2019unsupervised,sun2016return,pei2018multi}. In this work, we propose a simple and robust maximum mean discrepancy (MMD)-based alignment scheme \cite{pan2010domain,long2015learning}, and aim to achieve precise material identification under realistic constraints, including coarse momentum information and cross-domain deployment.

Building upon our previous coarse Z-level identification work \cite{wang2026transfer}, we propose a Coarse Momentum–Aware Domain Adaptation (CMADA) method for precise identification of nine specific materials. CMADA combines coarsely binned momentum cues with MMD-based unsupervised adaptation to improve both within-domain identification and cross-domain robustness. The main contributions are summarized as follows:

1. A coarse momentum-binning scheme is introduced to reduce the dependence on precise momentum measurements. The proposed scheme enables coarsely binned momentum information to achieve identification performance comparable to that obtained with fully precise momentum measurements. By combining 5-bin momentum scheme with a sample-statistics ment scheme, a material identification accuracy of 93.30\% is achieved.

2. Under a no–target-label setting, the proposed method improves cross-domain identification accuracy from 71.71\% (source-only training) to 89.00\%, effectively mitigating feature distribution misalignment in cross-domain scenarios.

3. The proposed method substantially reduces the hardware requirements of muon scattering-based material identification systems, as well as the data acquisition cost during deployment, thereby improving the feasibility of practical engineering implementation.

\section{\label{sec:Problem Setup and Data} Data Preparations}

\subsection{Coarse momentum encoding} 
Muon multiple Coulomb scattering depends on both material properties and the incident muon momentum. When momentum information is neglected, the resulting scattering-angle distribution deviates from an ideal Gaussian profile because low-momentum muons produce disproportionately large deflections. Consequently, a low-Z material traversed by a low-momentum muon may exhibit scattering characteristics similar to those of a high-Z material traversed by a high-momentum muon. This increases the overlap between the scattering-angle distributions of different material classes. However, in practical applications, precise muon momentum measurement is often impractical because of its cost and complexity \cite{hoepfner2025precise}. Thus, to improve the accuracy of muon-based material identification without relying on a high-precision momentum spectrometer \cite{priedhorsky2003detection}, coarse momentum information is incorporated through a momentum-integrated scattering angle $\widetilde{\theta}$ combined with momentum binning.

Based on the momentum dependence described in Eq.~\ref{eq:Eq.1}, a momentum-integrated scattering angle $\widetilde{\theta}$ is introduced. It is obtained by scaling the raw muon scattering angle using the ratio of the estimated muon momentum to a fixed reference momentum. This transformation maps scattering angles measured at different momenta to a common reference scale. It improves the consistency of the resulting $\widetilde{\theta}$ distributions and enhances material discriminability under limited-statistics conditions.

\begin{equation}
    \theta \approx \frac{13.6 \mathrm{MeV}}{\beta c p} \sqrt{\frac{L}{L_{0}}} \left[ 1 + 0.038 \ln \left( \frac{L}{L_{0}} \right) \right]
    \label{eq:Eq.1} 
\end{equation}

\begin{equation}
    \widetilde{\theta} = \theta \frac{\widetilde{p}} {p_{0}}
    \label{eq:momentum-weighted scattering angle}
\end{equation}
where $\theta$ denotes the raw scattering angle of an individual muon event, $\beta$ is the muon velocity normalized to the speed of light $c$, $p$ is the muon momentum, $L$ is the muon path length through the material, $L_0$ is the material radiation length, and $p_0$ is a fixed reference momentum selected as the mean muon momentum ($p_0 = 3000$MeV/$c$) \cite{yu2025improving}. 

\begin{figure}[htbp]
    \centering
    \includegraphics[width=0.65\textwidth]{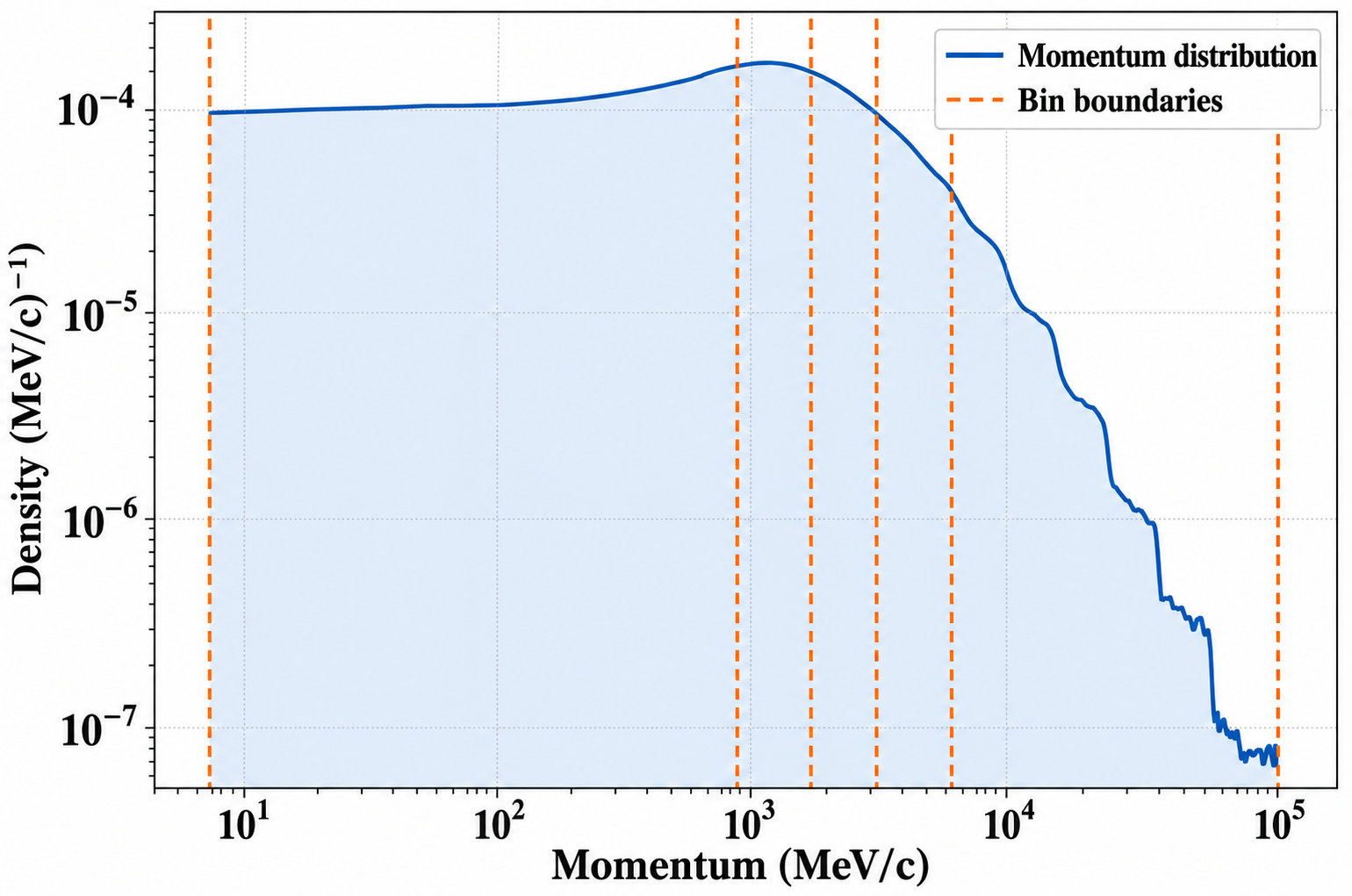}
    \caption{Momentum distribution and equal-frequency 5-bin scheme for cosmic-ray muon. The shaded regions denote momentum bins containing approximately equal numbers of muon events.}
    \label{fig:5bin}
\end{figure}

To obtain a coarse momentum estimate for each individual muon event, an equi-percentage momentum-binning scheme is adopted \cite{yu2025improving}. Specifically, the cosmic-ray muon momentum distribution is divided into $B$ bins and each bin contains approximately the same number of muon events, where $B$ denotes the number of momentum bins. A representative momentum value $\widetilde{p}$ is then assigned to all events within each bin. In this work, the median momentum of each bin is used as representative momentum value $\widetilde{p}$. The $\widetilde{p}$ is subsequently used to construct the momentum-integrated scattering angle $\widetilde{\theta}$. Figure~\ref{fig:5bin} illustrates the momentum distribution and the corresponding binning result for $B=5$. The curve represents the momentum distribution of cosmic-ray muons, while the shaded regions indicate the five momentum bins. The bin widths vary across the momentum range, becoming wider in sparsely populated regions and narrower in densely populated regions to maintain a same number of events in each bin.

\subsection{Simulation setups}

\begin{figure}[htbp]
    \centering
    \includegraphics[width=0.65\textwidth]{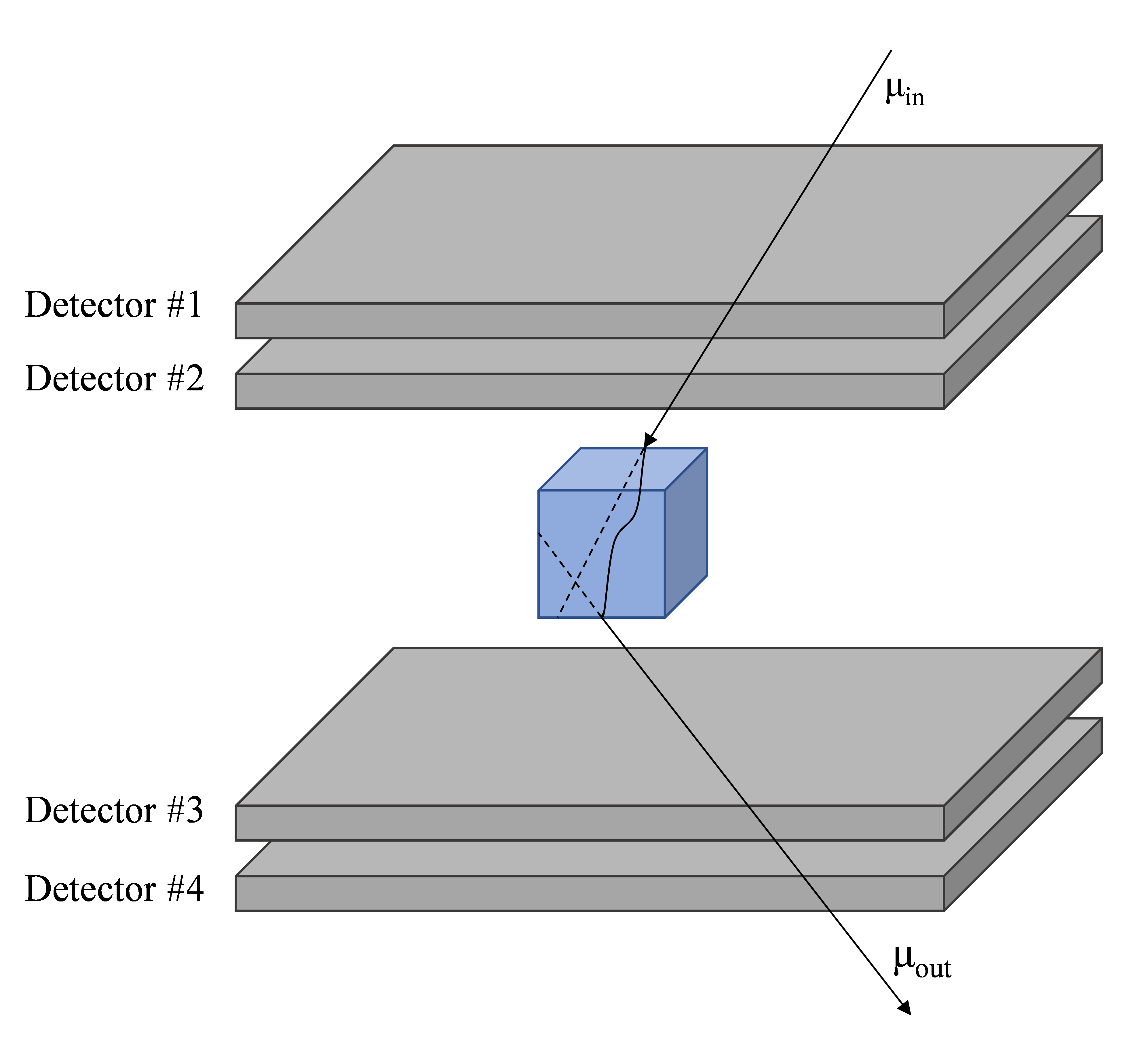}
    \caption{Illustration of the simulation model. A $10\,\text{cm} \times 10\,\text{cm} \times 10\,\text{cm}$ cubic target is positioned between two pairs of $30\,\text{cm} \times 30\,\text{cm}$ detector planes for muon scattering-angle reconstruction. The separations between the upstream and downstream detector pairs are 35 cm and 20 cm.}
    \label{fig:GEANT}
\end{figure}

Simulations were performed using Geant4 \cite{bury2025scattering}, which is widely used to model particle transport in matter. Cosmic-ray muons were generated from a planar source using the CRY library \cite{collaboration2003geant4}. The target is modeled as a $10\,\text{cm} \times 10\,\text{cm} \times 10\,\text{cm}$ cube. Muon trajectories are deflected by multiple Coulomb scattering, producing an angular deviation $\theta$ with an approximately Gaussian distribution \cite{allison2006geant4}. The scattering angle is measured by four $30\,\text{cm} \times 30\,\text{cm}$ detector planes installed upstream and downstream of the target, which are spaced 35 cm and 20 cm apart, respectively (see Fig.~\ref{fig:GEANT}).

For dataset construction, each muon traversal is treated as an independent event with recorded scattering angle $\theta$ and momentum $p$. For each material under different coating conditions, namely bare, Al-coated, and PE-coated, 500,000 events are simulated and stored to ensure sufficient statistical coverage. For the coated cases, the thickness of the Al or PE layer is $1\,\mathrm{cm}$.

Each sample consists of $N=500$ events and is represented as a $1\times500$ scattering-angle vector, where each element corresponds to a single muon event. To study cross-domain generalization, the detector settings are kept fixed while only the coating condition is varied. Domains are therefore defined according to the coating condition and share the same material label space. The bare condition serves as the labeled source domain, whereas the Al-coated and PE-coated conditions are treated as unlabeled target domains.

\section{Neural Network Models}

\subsection{Unsupervised domain adaptation with CMADA}

\begin{figure*}[htbp]
    \centering
    \includegraphics[width=0.95\textwidth]{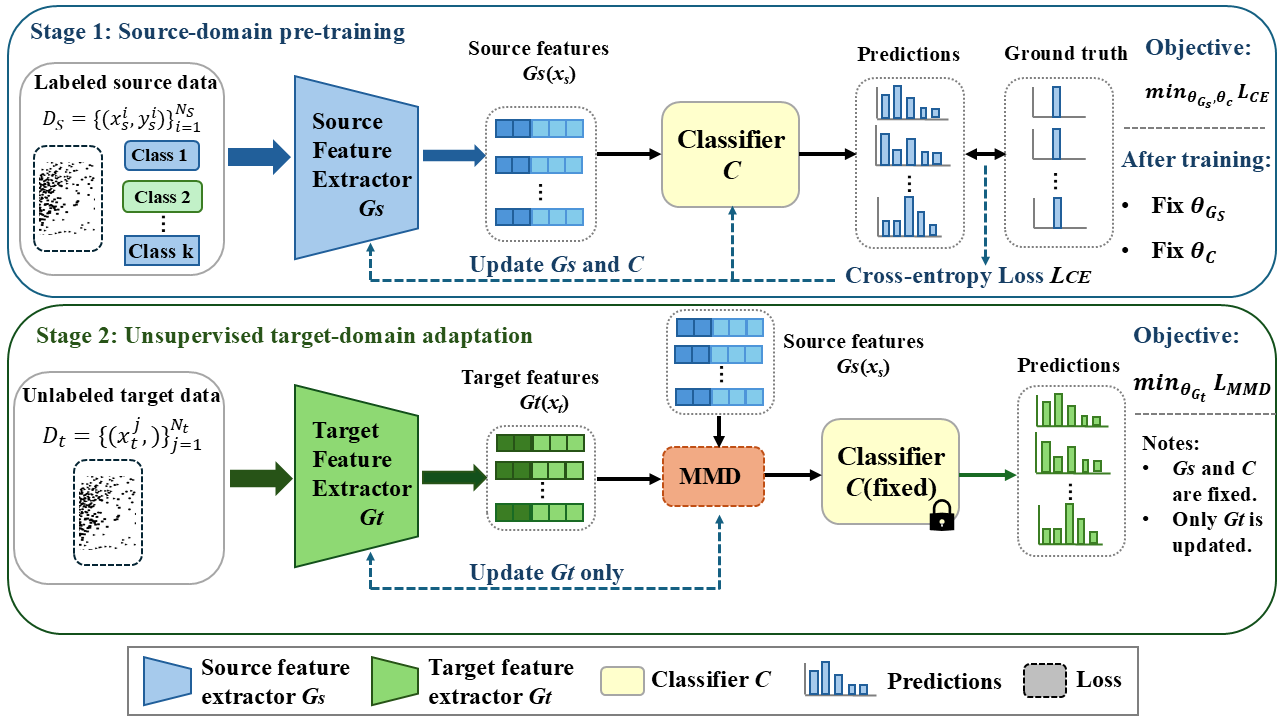}
    \caption{Process of CMADA. source domain pre-training: the source feature extractor and classifier are trained with source data, and gradient backpropagation is implemented via cross‑entropy loss. Unsupervised target domain adaptation stage: parameters of the source feature extractor are frozen, the target feature extractor is initialized, and MMD loss is used to align feature distributions between source and target domains for cross‑domain feature adaptation.}
    \label{fig:process_cmada}
\end{figure*}

To address the distribution mismatch between the source and target domains, a Coarse Momentum-Aware Domain Adaptation (CMADA) framework is proposed based on momentum-integrated features generated through coarse momentum binning. As illustrated in Fig.~\ref{fig:process_cmada}, the framework consists of a source feature extractor $G_s$, a target feature extractor $G_t$, and a shared classifier $C$. The source branch is trained using labeled source domain samples to establish a discriminative feature space for material identification, while the target branch is optimized to align unlabeled target domain features with the source domain representation space. The classifier is shared by both branches to ensure category consistency during adaptation.

The target feature extractor $G_t$ adopts the same architecture as $G_s$ and is initialized using the pretrained parameters of the source branch. During adaptation, only $G_t$ is updated, allowing the target domain features to gradually align with the source domain distribution while preserving the discriminative feature structure learned from the source domain.

The shared classifier $C$ maps the extracted features into material categories. After supervised training on the source domain, the classifier parameters are fixed throughout the adaptation stage to provide a stable decision boundary for unlabeled target domain samples. This strategy prevents the identification boundary from drifting during optimization and encourages the aligned target domain features to preserve category-level discriminability while reducing marginal distribution discrepancy between domains.

The training procedure consists of two stages. In the first stage, $G_s$ and $C$ are trained using the labeled source dataset $\mathcal{D}_s=\left\{\left(x_s^{(i)},\,y_s^{(i)}\right)\right\}_{i=1}^{N_s}$. The network is optimized using the cross-entropy (CE) loss to learn discriminative material identification from momentum-integrated scattering features. Through supervised training, the extracted features become less sensitive to statistical fluctuations and momentum uncertainty. The CE loss is defined as follows:

\begin{equation}
   \min_{\theta_{G_s}, \theta_C} \mathcal{L}_{CE} = - \mathbb{E}_{(x_s, y_s) \sim \mathcal{D}_s} \sum_{k=1}^{K} \mathbb{I}_{[y_s,k]} \log C(G_s(x_s)) 
\end{equation}
where $\theta_{G_s}$ and $\theta_C$ denote the parameters of $G_s$ and $C$, respectively \cite{lin2025simulation}. $K=9$ is the number of material classes, and $\mathbb{I}$ is the indicator function. An early stopping scheme based on a source validation set is employed to improve generalization. After this supervised source domain training step, both $\theta_{G_s}$ and $\theta_C$ are fixed.

Subsequently, unsupervised target domain adaptation is performed by optimizing only $G_t$. $G_t$ is initialized using the pretrained parameters of $G_s$, while both $G_s$ and $C$ remain fixed throughout the adaptation process. In this design, the source domain feature space produced by $G_s$ serves as a fixed reference, and $G_t$ is optimized to align target domain features with the source domain distribution by minimizing the maximum mean discrepancy (MMD) between $G_s(X_s)$ and $G_t(X_t)$. After feature alignment, $C$ can be used directly for target domain material identification without requiring target domain labels. The adaptation objective is defined as:

\begin{equation}
    \min_{\theta_{G_t}} \mathcal{L}_{MMD} = \left\| \frac{1}{N_s} \sum_{i=1}^{N_s} \phi(G_s(x_s^i)) - \frac{1}{N_t} \sum_{j=1}^{N_t} \phi(G_t(x_t^j)) \right\|_{\mathcal{H}}^2
\end{equation}
where $N_s$ and $N_t$ denote the number of source domain and target domain samples used for MMD estimation. $\phi(\cdot)$ denotes the kernel-induced feature mapping, implemented using a Gaussian RBF kernel in this work. $\mathcal{H}$ denotes the Reproducing Kernel Hilbert Space (RKHS) \cite{pan2010domain,gretton2012kernel}.

The core idea of MMD is straightforward: if two distributions are identical, the expectations of samples drawn from them are equal for any function within a given function space. Since MMD does not require labels from the target domain, it can effectively reduce domain discrepancies and improve the model performance on the target domain. Commonly used kernel functions include the linear kernel, polynomial kernel, and Gaussian (RBF ) kernel. In this work, the Gaussian kernel is adopted. Its effective range is controlled by the bandwidth parameter $\sigma$, which determines the sensitivity of MMD to both local and global distribution discrepancies. An excessively small $\sigma$ tends to overemphasize local details, leading to unstable MMD values and potential overfitting. In contrast, an excessively large $\sigma$ results in excessive smoothing, making the kernel less sensitive to distribution differences and causing the MMD value to approach zero. Therefore, this work employs the median heuristic strategy to determine $\sigma$. Specifically, the source and target datasets are first merged, and the squared Euclidean distances between all sample pairs are computed. Let $\mathrm{med}^2$ denote the median of these squared distances. The bandwidth is then defined as

\begin{equation}
\sigma = \frac{\sqrt{\mathrm{med}^2}}{2}.
\end{equation}

MMD is employed to align the feature distributions of the source and target domains in the latent feature space. Although MMD effectively reduces the discrepancy between the source and target domains during training, the absence of target domain labels still makes model selection challenging in unsupervised domain adaptation. To address this issue, Information Maximization (IM) is adopted as an unsupervised criterion for checkpoint selection. This criterion is designed to favor models that produce confident predictions for individual target samples while maintaining a diverse class distribution across the target dataset. The IM objective function $\mathcal{J}_{IM}$ is defined as the difference between the conditional entropy and the marginal entropy \cite{liang2020we}. For the target domain dataset $\mathcal{D}_t = \{x_t^j\}_{j=1}^{N_t}$, where $N_t$ denotes the number of target samples, the objective is formulated as follows:

\begin{equation}
    \mathcal{J}_{IM} = \mathcal{H}_{cond} - \mathcal{H}_{marg}
\end{equation}

\begin{equation}
    \mathcal{H}_{cond} = \frac{1}{N_t} \sum_{j=1}^{N_t} \sum_{k=1}^{K} -p(k|x_t^j) \log p(k|x_t^j)
\end{equation}

\begin{equation}
    \mathcal{H}_{marg} = \sum_{k=1}^{K} -\bar{p}_k \log \bar{p}_k, 
    \quad \text{with} \quad 
    \bar{p}_k = \frac{1}{N_t} \sum_{j=1}^{N_t} p(k|x_t^j)
\end{equation}
where $p(k|x_t^j)$ denotes the predicted probability that the target sample $x_t^j$ belongs to class $k$, $K$ is the number of material classes, and $\bar{p}_k$ represents the average predicted probability assigned to class $k$ over the entire target domain dataset.

The information maximization objective consists of two complementary terms: the conditional entropy term $\mathcal{H}_{cond}$ and the marginal entropy term $\mathcal{H}_{marg}$. The conditional entropy term $\mathcal{H}_{cond}$ reduces the uncertainty of individual predictions, thereby encouraging the model to produce confident identification outputs. In contrast, the marginal entropy term $\mathcal{H}_{marg}$ maximizes the entropy of the batch-averaged prediction distribution $\bar{p}$, encouraging diversity across predicted categories and preventing the model from collapsing to trivial single-class prediction. The combination of these two terms balances prediction confidence and category diversity during adaptation. Notably, $\mathcal{J}_{IM}$ is used only as an unsupervised validation criterion for checkpoint selection, rather than as a training objective for backpropagation. During domain adaptation, the MMD loss $\mathcal{L}_{MMD}$ usually decreases rapidly at the early stage and then reaches a relatively stable stage. After this stage is reached, $\mathcal{J}_{IM}$ is evaluated on the target dataset for each checkpoint, and the checkpoint with the minimum $\mathcal{J}_{IM}$ is selected as the optimal checkpoint, since it provides a balance between prediction confidence and class diversity.

\begin{table*}[t]
\centering
\caption{Hyperparameter settings used for the model.}
\label{tab:model_parameters}
\begin{tabular}{@{} l c p{6cm} @{}}
\toprule
\textbf{Parameter} & \textbf{Value} & \textbf{Description} \\
\midrule
$\mathrm{input\_dim}_{f}$      & $n$       & Feature dimension of the input data \\
$\mathrm{hidden\_dim}_{f,1}$   & 128       & Number of nodes in the first hidden layer of the feature extractor \\
$\mathrm{BatchNorm1d}$         & 128       & Number of features in the batch normalization layer \\
$\mathrm{hidden\_dim}_{f,2}$   & 128       & Number of nodes in the second hidden layer of the feature extractor \\
$\mathrm{hidden\_dim}_{c,1}$   & 128       & Number of nodes in the first hidden layer of the classifier \\
$\mathrm{hidden\_dim}_{c,2}$   & 64        & Number of nodes in the second hidden layer of the classifier \\
$\mathrm{output\_dim}_{c}$     & 9         & Number of classes in the output layer \\
$\mathrm{hidden\_dim}_{d,1}$   & 128       & Number of nodes in the first hidden layer of the domain discriminator \\
$\mathrm{hidden\_dim}_{d,2}$   & 128       & Number of nodes in the second hidden layer of the domain discriminator \\
$\mathrm{output\_dim}_{d}$     & 1         & Output dimension of the domain discriminator \\
$\mathrm{tt\_ratio}$           & 7{:}3     & Ratio of training to test samples \\
$\mathrm{batch\_size}$         & 256       & Number of samples per training batch \\
$\mathrm{epochs}$              & 100       & Total number of training epochs \\
$\mathrm{lr}_{f}$              & $1\times10^{-4}$ & Learning rate for the feature extractor \\
$\mathrm{lr}_{c}$              & $1\times10^{-4}$ & Learning rate for the classifier \\
$\mathrm{lr}_{d}$              & $1\times10^{-4}$ & Learning rate for the domain discriminator \\
$\mathrm{lr}_{mmd}$            & $5\times10^{-4}$ & Learning rate for the MMD module \\
$\mathrm{lr}_{pretrain}$       & $1\times10^{-4}$ & Learning rate during pre-training \\
$\mathrm{kernel\_num}$         & 5           & Number of Gaussian kernels \\
$\mathrm{bandwidth}$           & 2.0         & Bandwidth of the Gaussian kernel \\

\bottomrule
\end{tabular}
\end{table*}

\subsection{Model settings}

To evaluate the effectiveness of the proposed model, comparative studies are conducted using the DANN and source-only baselines. A strict unsupervised domain adaptation protocol is adopted, in which the target domain dataset is divided into an unlabeled training set and a test set. The unlabeled target training set is used exclusively for domain alignment, without access to any target labels, while the target test labels are reserved solely for the final performance evaluation. The following adaptation methods and baseline models are implemented and compared:

1. Source-only (No Adaptation): This baseline quantifies the raw performance degradation caused by distribution shifts. We first train $G$ and $C$ to convergence on the labeled Source domain with Cross-Entropy loss. This pre-trained model is then directly applied to the target test set without any further tuning or adaptation.

2. DANN \cite{ganin2016domain} (Adversarial Baseline): The Domain-Adversarial Neural Network serves as a robust baseline to evaluate the stability of the proposed MMD-based alignment. The training involves a minimax game where a domain discriminator attempts to distinguish between source and target features, while the feature extractor is trained to deceive the discriminator via a Gradient Reversal Layer (GRL), thereby learning domain-invariant representations.

3. CMADA (Ours): Firstly, $G_t$ is initialized with the weights of the pre-trained  $G_s$. Then, the optimization minimizes the MMD distance between source and target feature distributions. In each training iteration, the network receives a mixed batch containing labeled source samples and unlabeled target samples. Notably, $G_s$ and $C$ are frozen, and only $G_t$ is updated. Finally, during the inference stage, this adapted target extractor is combined with the fixed classifier to predict material classes on the strictly reserved target test set.

To ensure a fair and comprehensive comparison, a unified protocol was adopted. All comparative methods, including the Source-only baseline, DANN, and the proposed CMADA, utilize the identical network backbone and input feature dimension . This constraint ensures that any observed performance improvements are attributable solely to the effectiveness of the domain adaptation method rather than improvements in the model capacity or feature engineering. Additionally, the DANN baseline was adapted to follow the same decoupled two-stage framework as CMADA.

Since the adaptation stage is independent of the identification task in both CMADA and DANN, the models are optimized solely through domain alignment objectives, namely $L_{\mathrm{MMD}}$ for CMADA and $L_{\mathrm{adv}}$ for the modified DANN. For both methods, the optimal checkpoint is selected according to the IM score evaluated on the target training set. Table~\ref{tab:model_parameters} summarizes the hyperparameter settings, including the learning rates of different modules, network architectures, and training epochs. All models were trained using the Adam optimizer with a batch size of 256. The algorithms were implemented in PyTorch and trained on a single NVIDIA RTX 3090 GPU.

\subsection{Evaluation metrics}

In this paper, the performance of the proposed framework is evaluated using three complementary metrics: feature separability, overall identification performance, and class-wise prediction details.

Silhouette Coefficient (SC): Feature separability is quantified using the silhouette coefficient. This metric evaluates the learned feature by measuring both intra-class compactness and inter-class separation. A higher SC value indicates that samples from the same material class form more compact clusters and are better separated from samples of other classes. For a given sample $x_i$ belonging to class $C_I$, the silhouette score $s(i)$ is defined as:

\begin{equation}
    s(i) = \frac{b(i) - a(i)}{\max\{a(i), b(i)\}}
\end{equation}

\begin{equation}
    a(i) = \frac{1}{|C_I| - 1} \sum_{x_j \in C_I, j \neq i} d(x_i, x_j)
\end{equation}

\begin{equation}
    b(i) = \min_{J \neq I} \left( \frac{1}{|C_J|} \sum_{x_j \in C_J} d(x_i, x_j) \right)
\end{equation}
where $a(i)$ and $b(i)$ denote the mean intra-cluster and nearest-cluster distances, measuring cohesion and separation, respectively \cite{ROUSSEEUW198753}. The distance $d(\cdot, \cdot)$ is Euclidean in feature space. The overall SC is the dataset average of $s(i)$, where values approaching 1 indicate compact and well-separated clusters, reflecting strong feature discriminability.

Identification Accuracy: Overall performance is
evaluated using classification accuracy. The overall accuracy is calculated as the ratio of correctly predicted samples to the total number of samples $N$:

\begin{equation}
    \text{Accuracy} = \frac{1}{N} \sum_{i=1}^{N} \mathbb{I}(\hat{y}_i, y_i) \times 100\%
\end{equation}
where $y_i$ and $\hat{y}_i$ denote the ground-truth label and the predicted label of the i-th sample, respectively. $\mathbb{I}(\cdot)$ is the indicator function, which equals 1 if $\hat{y}_i = y_i$ and 0 otherwise.

Principal Component Analysis (PCA) Visualization: PCA projects high-dimensional features into a low-dimensional space (typically 2D) to qualitatively assess (i) class separability and (ii) domain alignment by visualizing the overlap between source and target distributions before and after adaptation.

\subsection{Scanning modes}

In practical applications, data from a single scan are often insufficient for stable estimation due to statistical fluctuations in muon events, while increasing the number of scans directly leads to higher acquisition time and reduced throughput. This introduces a trade-off between measurement efficiency and statistical reliability. To address this issue, we design two scanning modes that offer different balances between speed and precision.

\begin{figure*}[htbp]
    \centering
    \includegraphics[width=0.95\linewidth]{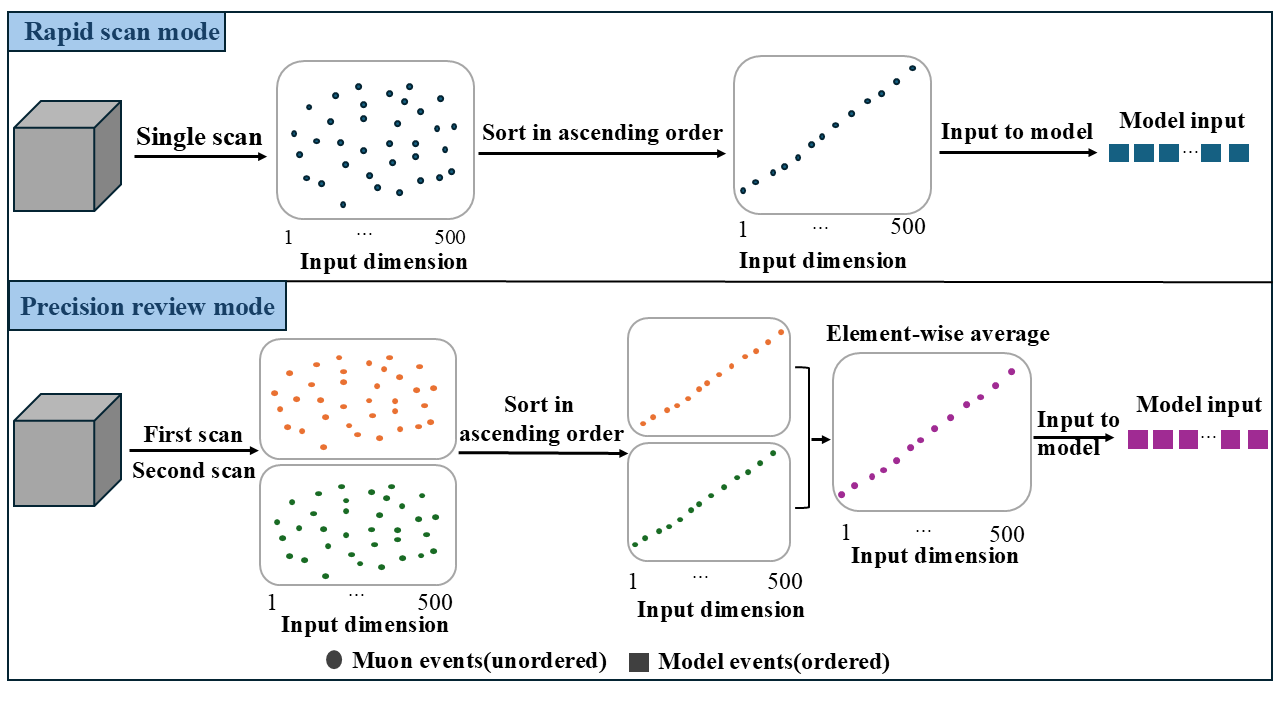}
    \caption{Schematic flowcharts of the Rapid scan mode and the proposed Precision review mode, respectively.}
    \label{fig:Fig_mode}
\end{figure*}

In the rapid scan mode used in previous analyses, each sample is constructed from 500 muon events collected within a single scan and represented as a 500-dimensional scattering-angle vector. This setting prioritizes acquisition efficiency but is more sensitive to event-level fluctuations. In the precision review mode, two independent scans are performed sequentially on the same object, each containing 500 muon events. The resulting scattering-angle vectors are averaged, which effectively suppresses statistical noise and improves estimation stability. Since the input dimensionality remains unchanged, the previously trained model can be directly applied without modifying the network architecture or retraining the classifier. A schematic comparison of the two modes is shown in Fig.~\ref{fig:Fig_mode}.

\section{\label{sec:Results_discussion}Results and discussion}
The effectiveness of the proposed framework is evaluated in this section using Geant4-based simulation data, which is divided into three parts. First, the influence of incorporating momentum information on source domain material feature separability is analyzed. Then, source domain material identification performance under the broad momentum distribution of cosmic-ray muons is evaluated using momentum-binning schemes, and the effect of the precision review mode is also analyzed. Finally, the cross-domain generalization ability of the proposed unsupervised domain adaptation method is evaluated.

\subsection{Influence of momentum Information in the source domain}
\label{results1}

\begin{figure}[htbp]
    \centering
    \begin{subfigure}[t]{0.45\textwidth}
        \centering
        \includegraphics[width=\linewidth]{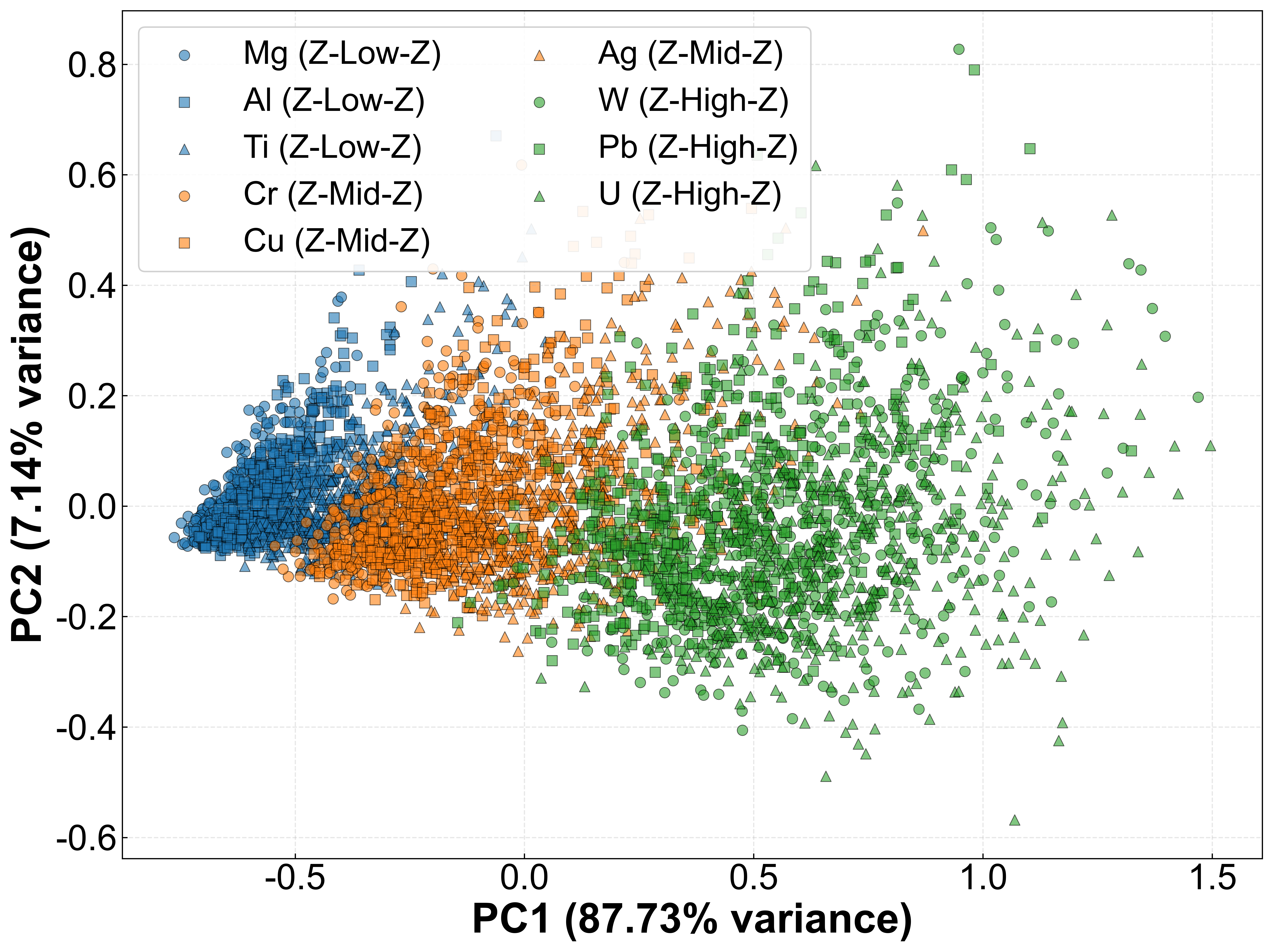}
        \caption{PCA visualization of raw scattering angle features in the source domain. Significant overlap is observed among different material groups.}
        \label{fig:pca_angle}
    \end{subfigure}
    \hfill
    \begin{subfigure}[t]{0.45\textwidth}
        \centering
        \includegraphics[width=\linewidth]{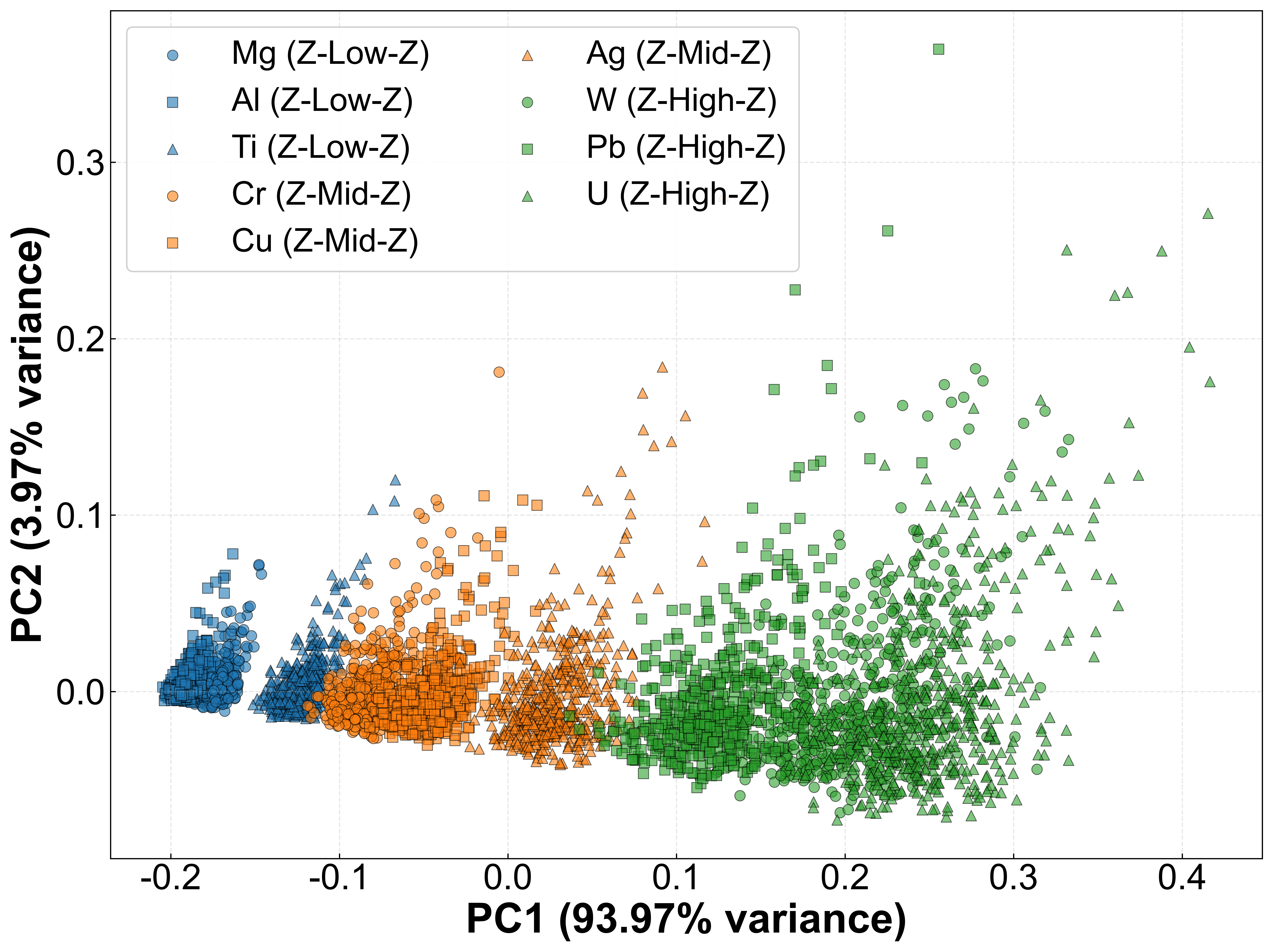}
        \caption{PCA visualization of processed scattering angle features in the source domain. The introduction of momentum information leads to improved the separability of different materials within each Z group.}
        \label{fig:pca_p}
    \end{subfigure}
    \caption{PCA visualization for raw and processed scattering angles in the source domain.}
    \label{fig:pca_comparison}
\end{figure}

To illustrate the importance of momentum information in muon-based material identification, the feature-space structures under two representative cases, i.e., without momentum information and with precise momentum information, are compared. Fig.~\ref{fig:pca_comparison} presents the PCA projections obtained from raw scattering angles and processed scattering angles. Colors and marker shapes denote different material groups and material types.

In Fig.~\ref{fig:pca_comparison}(a), materials from different Z groups are well separated, while those within the same group overlap heavily. This shows raw scattering angles enable coarse Z identification but not precise material identification. This limitation mainly arises from the unknown muon momentum, which broadens the scattering-angle distributions and weakens material-dependent differences. After momentum information is introduced in Fig.~\ref{fig:pca_comparison}(b), the overlap between Z groups is further reduced and the separability within each group is significantly improved. In particular, Ti, Ag, and Pb form relatively independent clusters, while slight overlap remains between Mg/Al, Cr/Cu, and U/W. These results indicate that momentum information effectively suppresses momentum-induced variations in the scattering-angle distributions and enhances material discriminability, whereas the remaining overlap mainly reflects the intrinsic similarity of the scattering behaviour of certain materials.

The PCA results indicate that incorporating momentum information significantly improves material separability in the feature space by reducing the momentum-induced broadening of scattering-angle distributions. The remaining overlap is mainly associated with the intrinsic similarity of the scattering behaviour of certain materials, particularly within the high-Z group.

\textbf{}

\begin{table}[htbp]
  \centering
  \caption{Class-wise identification accuracy (\%) of the source-only model under representative momentum binning schemes (single sampling). The 5-bin momentum binning scheme achieves a comparable level of accuracy to that of the precise momentum scheme.}
  \label{tab: accuracy_comparison}
  \begin{tabular}{lcccc}
    \toprule
    Material & 0-bin & 3-bin & \textbf{5-bin (Baseline)} & Precise (Bound) \\
    \midrule
    Mg (Low-Z)  & 69.67 & 93.33 & 95.33 & 97.33 \\
    Al (Low-Z)  & 74.67 & 95.00 & 97.33 & 98.33 \\
    Ti (Low-Z)  & 71.67 & 95.00 & 96.00 & 100.00 \\
    \addlinespace
    Cr (Mid-Z)  & 54.33 & 85.67 & 92.00 & 96.67 \\
    Cu (Mid-Z)  & 63.33 & 87.00 & 93.00 & 92.00 \\
    Ag (Mid-Z)  & 65.00 & 95.67 & 99.33 & 98.67 \\
    \addlinespace
    W (High-Z)  & 36.33 & 60.33 & 57.33 & 68.67 \\
    Pb (High-Z) & 58.67 & 89.00 & 93.33 & 97.00 \\
    U (High-Z)  & 65.67 & 71.33 & 82.00 & 81.33 \\
    \midrule
    \textbf{Overall} & 62.15 & 85.81 & \textbf{89.52} & 92.22 \\
    \bottomrule
  \end{tabular}
\end{table}

\subsection{source domain material identification with momentum-binning schemes}
\label{results2}

The above results show that material separability can be significantly improved when precise momentum information is introduced. However, in practical applications, precise muon momentum measurement is usually not feasible. For this purpose, SC and identification accuracy are evaluated under different momentum-binning schemes, as shown in Fig.~\ref{fig:momentum_granularity}. Figure~\ref{fig:momentum_granularity}(a) and (b) present SC and the identification accuracy under different momentum-binning schemes. Here, the identification accuracy is used as a complementary metric for material separability. For each momentum-binning scheme, the classifier is trained on the source domain training set, and the reported identification accuracy is obtained from the source domain test set. 

\begin{figure}[htbp]
    \centering
    \includegraphics[width=1\linewidth]{Fig5_SC.png}
    \caption{Impact of momentum-binning scheme on feature separability and identification performance across three material groups in the source domain. (a) SC for feature separability and (b) Identification accuracy under different momentum binning schemes.}
    \label{fig:momentum_granularity}
\end{figure}

As shown in Fig.~\ref{fig:momentum_granularity}(a), both the SC and the identification accuracy improve as the momentum binning becomes finer. Without momentum information, the raw scattering-angle features present the weakest separability, with the lowest SC values observed across all material groups and negative SC values appearing for the mid-Z and high-Z groups. Correspondingly, the overall identification accuracy is also the lowest (62.15\%). After momentum information is introduced, both metrics improve significantly, particularly from the no-momentum case to the 3-bin and 5-bin schemes. The overall SC increases from approximately 0.02 to 0.12 and 0.20, while the overall identification accuracy rises to 85.81\% and 89.52\%. Beyond the 5-bin scheme, the improvement gradually saturates. For the 7-, 9-, and 11-bin schemes, the identification accuracy remains within 89\%--91\%, and the SC improvement over the 5-bin scheme is minor. Although the precise-momentum case achieves the best performance, with an SC of approximately 0.30 and an accuracy of 92.22\%, its advantage over the 5-bin and 7-bin schemes remains limited. Since finer momentum binning requires more accurate momentum estimation and introduces additional implementation complexity, the 5-bin scheme is adopted as the baseline setting in the following analysis. Based on this setting, further improvements in practical identification performance are subsequently investigated.

To further evaluate material-level identification performance, the class-wise accuracy under different momentum schemes are summarized in Table~\ref{tab: accuracy_comparison}. Under the 5-bin scheme, all low-Z and mid-Z materials achieve accuracy above 90\%, indicating that coarse momentum information is already sufficient for reliable identification in these groups. In contrast, the high-Z group remains significantly more challenging. Although Pb reaches an accuracy of 93.33\%, the accuracy of U and W are only 82.00\% and 57.33\%. Even in the precise-momentum case, the accuracy of W increases only to 68.67\%, which is still substantially lower than that of the low-Z and mid-Z materials. This trend is consistent with the group analysis, where the low-Z group achieves the highest SC values and identification accuracy, followed by the mid-Z group, while the high-Z group remains the most difficult. From these results, the remaining difficulty mainly originates from the intrinsic similarity of the scattering behaviour of high-Z materials rather than from insufficient momentum information.

\begin{figure}[htbp]
    \centering
    \includegraphics[width=0.55\linewidth]{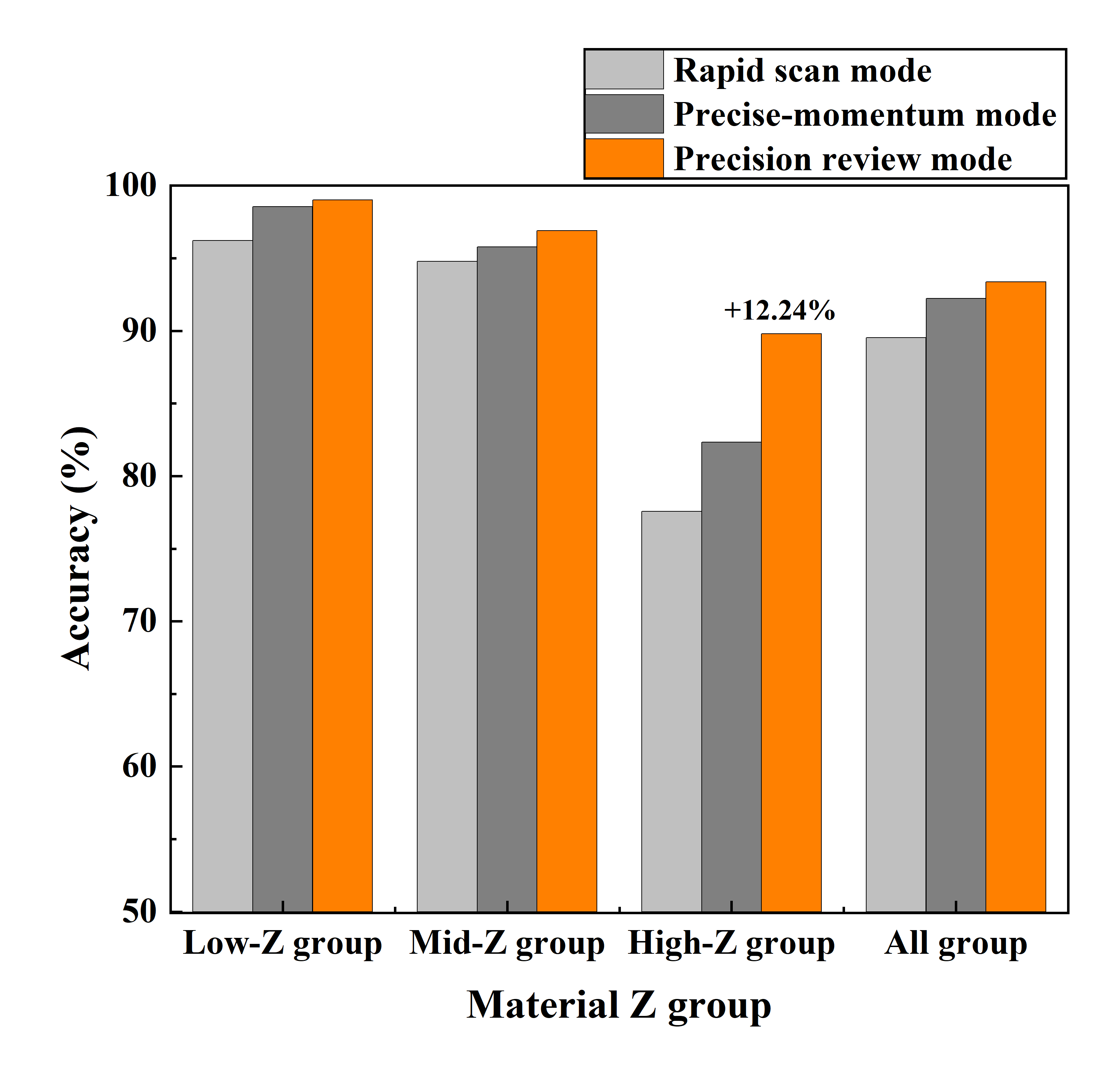}
    \caption{Performance comparison of material identification modes across different Z-groups in the source domain. The proposed mode achieves consistent accuracy improvements over both the baseline and precise-momentum mode, with a particularly significant improvement of 12.24\% accuracy gain over the baseline in the challenging high-Z group.}
    \label{fig:comparison_strategies}
\end{figure}

Figure~\ref{fig:comparison_strategies} compares three
identification modes: the 5-bin rapid scan, the 5-bin precision review, and the precise-momentum mode. The 5-bin precision review mode achieves the best performance across all material groups. The improvement is relatively limited for the low-Z and mid-Z groups, where the identification accuracy is already high in the rapid scan mode. In contrast, the high-Z group shows a much larger improvement, with the accuracy increasing from 77.67\% to 89.91\%. The overall accuracy of the 5-bin precision review mode reaches 93.37\%, slightly higher than the 92.22\% achieved by the precise-momentum mode. Although the latter relies on precise momentum measurements, the 5-bin precision review mode achieves comparable or better performance with only two scans and a simpler processing pipeline. Considering the trade-off between measurement efficiency and statistical reliability, the 5-bin precision review mode is adopted in the following experiments.

\subsection{Cross-domain material identification with unsupervised domain adaptation}
\label{results3}

The above results show that the setting consisting of the 5-bin momentum scheme and precision review mode achieve reliable material identification in the source domain. In this section, the cross-domain material identification performance under this setting is further evaluated. The models trained on the source domain are transferred to the target domain under a fully unsupervised setting. Three methods are compared: the Source-only model without adaptation, the adversarial baseline DANN, and the proposed CMADA.

It should be noted that different coating materials introduce different levels of domain shift. Due to its relatively small thickness and weak scattering effect, the PE-coated target domain exhibits only limited deviation from the source domain. In contrast, the Al-coated target domain produces a larger distribution shift because of its stronger scattering contribution, making the cross-domain identification task more challenging and closer to practical application scenarios. Therefore, the following discussion focuses on the Al-coated target domain.

\begin{table}[htbp]
\centering
\caption{Identification accuracy (\%) of different methods in the Al-coated target domain.
The accuracy are calculated from the corresponding confusion matrices, with 300 samples for each material.}
\label{tab:accuracy_from_cm}
\begin{tabular}{c|c|ccccccccc}
\hline
\multirow{2}{*}{Method}
& \multirow{2}{*}{\makecell{Overall\\Accuracy (\%)}}
& \multicolumn{9}{c}{Class-wise Identification Accuracy (\%)} \\
& & Mg & Al & Ti & Cr & Cu & Ag & W & Pb & U \\
\hline
Source-only
& 71.74
& \textbf{99.67}
& 0.00
& \textbf{99.67}
& 93.00
& 63.67
& 83.67
& 67.67
& 88.67
& 49.67 \\

DANN
& 83.26
& 98.67
& 40.67
& 97.00
& \textbf{96.33}
& 91.67
& 98.33
& 73.00
& 91.00
& 62.67 \\

CMADA
& \textbf{89.00}
& 70.00
& \textbf{98.33}
& 98.00
& 87.33
& \textbf{98.67}
& \textbf{100.00}
& \textbf{80.67}
& \textbf{95.33}
& \textbf{72.67} \\
\hline
\end{tabular}
\end{table}

Table~\ref{tab:accuracy_from_cm} summarizes each material and overall identification accuracy of the three methods. When the Source-only model is directly applied to the target domain, the overall accuracy decreases from 93.37\% in the source domain to 71.74\%, indicating that the representation learned from the source domain is strongly affected by the target domain shift and cannot be directly transferred to the target domain. After domain adaptation is introduced, the performance improves significantly: DANN increases the overall accuracy to 83.26\%, while CMADA further improves it to 89.00\%, demonstrating a stronger capability in reducing the discrepancy between the source and target domains.

\begin{figure}[!htbp]
    \centering
    \includegraphics[width=1\linewidth]{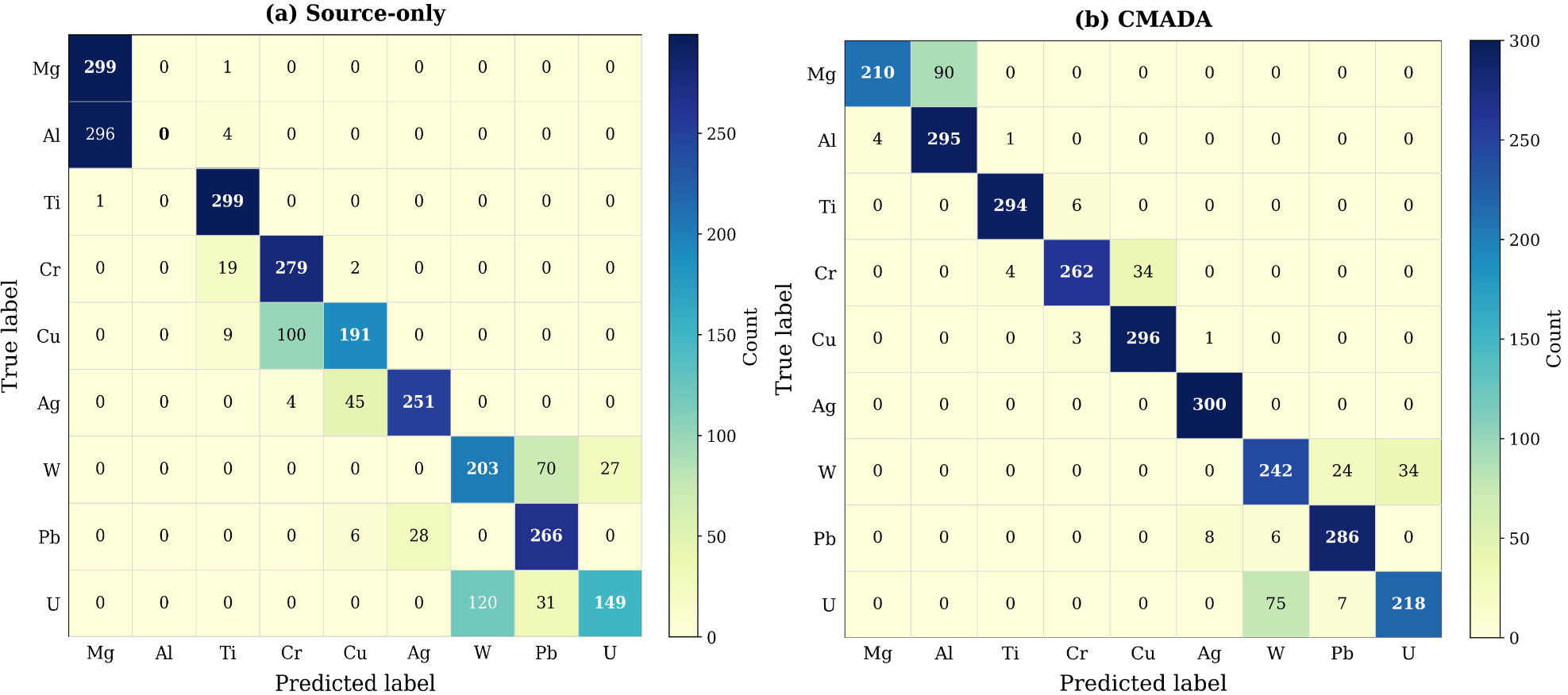}
    \caption{Confusion matrices of the Source-only model and the proposed CMADA in the Al target domain. Rows denote the true labels and columns denote the predicted labels.}
    \label{fig:confusion_matrix}
\end{figure}

Fig.~\ref{fig:confusion_matrix} further illustrates the prediction patterns for Source-only model and CMADA model. Since the prediction pattern of the DANN model is similar to that of the CMADA model, it is not shown here for brevity. Class-wise results and confusion matrices follow the same trend. The Source-only model performs poorly on multiple materials: accuracy drops to 0.00\% for Al and 49.67\% for U. Al samples are mostly identified as Mg, and substantial confusion is observed among high-Z materials, including W, Pb, and U. CMADA greatly improves Al recognition and reduces confusion in low- and mid-Z groups. The accuracy of Al and U rises to 98.33\% and 72.67\%, with Cu, Ag, W and Pb also achieving better results. Still, occasional confusion between W and U can still be observed. This reveals domain adaptation mitigates distribution mismatch between domains, yet cannot fully resolve identification challenges for materials with similar scattering characteristics.

\begin{figure}[!htbp]
    \centering
    \begin{subfigure}[t]{0.45\textwidth}
        \centering
        \includegraphics[height=5.5cm]{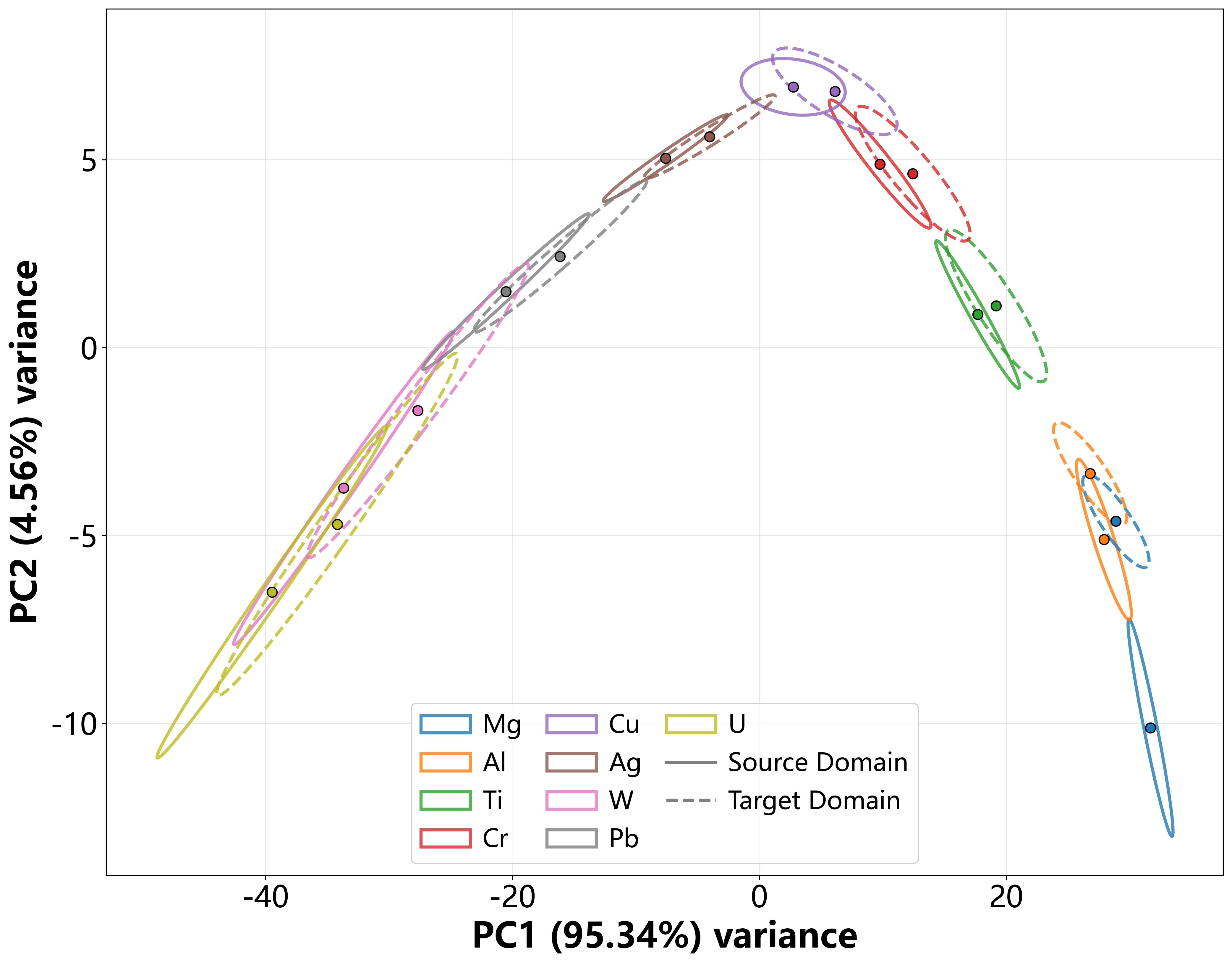}
        
        \caption{PCA visualization of Source-only model. A noticeable distribution discrepancy exists between the source domain and target domain features for multiple materials, indicating limited cross-domain generalization capability when no domain adaptation scheme is applied.}
        \label{fig:pca_source}
    \end{subfigure}
    \hfill
    \begin{subfigure}[t]{0.45\textwidth}
        \centering
        \includegraphics[height=5.5cm]{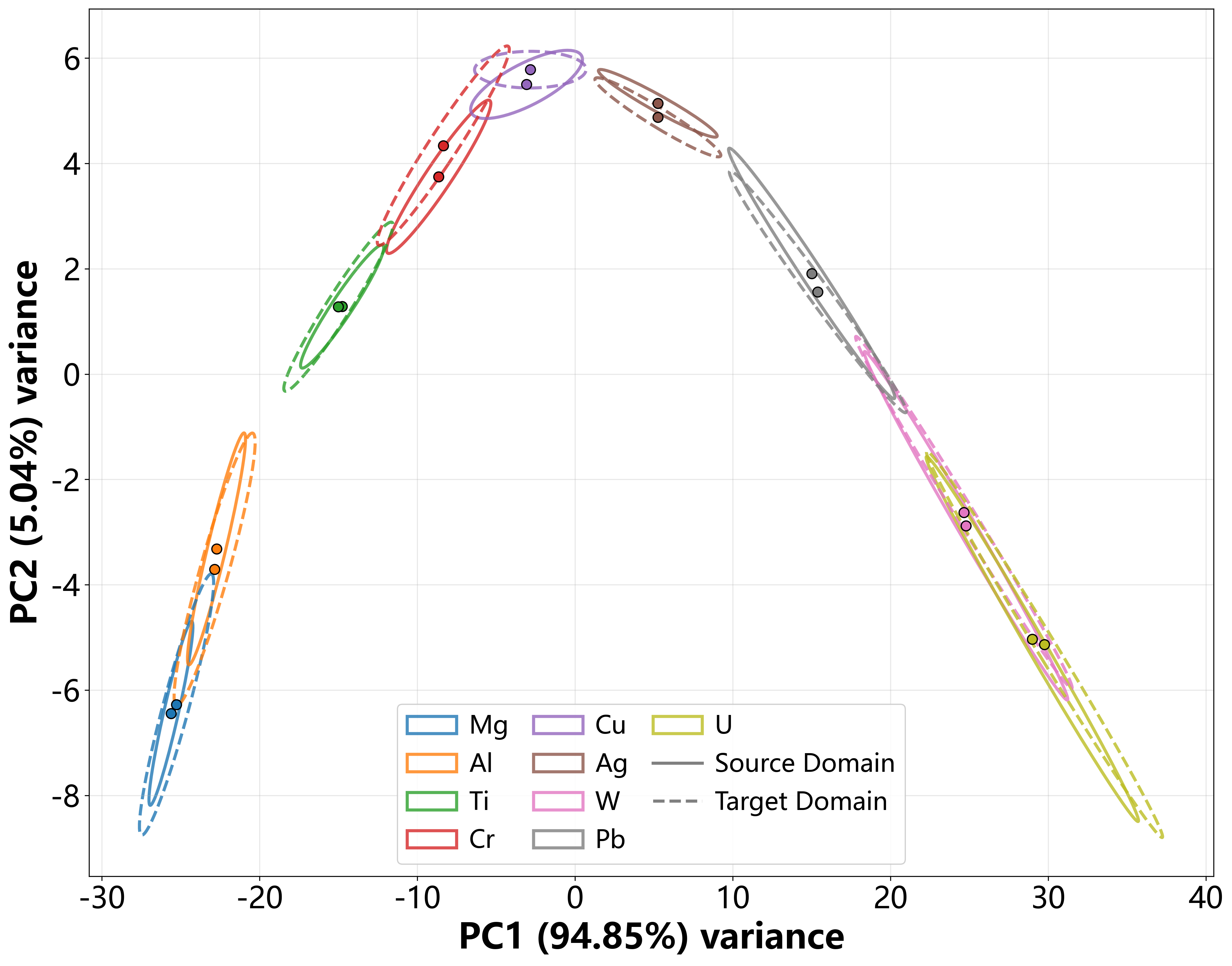}
        \caption{PCA visualization of  CMADA model. After domain adaptation, the feature distributions of the source and target domains become substantially better aligned for each material category, demonstrating that the proposed framework effectively mitigates the domain shift between simulated and measured data.}
        \label{fig:kde_p}
    \end{subfigure}
    \caption{Comparison of PCA visualizations of learned feature representations. }
    \label{fig:pca_cmada}
\end{figure}

Fig.~\ref{fig:pca_cmada} shows PCA visualizations before and after adaptation for Source-only model and CMADA model. Since the PCA visualizations of the DANN model is similar to that of the CMADA model, it is not shown here for brevity. In the Source-only model, a clear distribution mismatch exists between the source and target domains. In particular, the target domain Mg features shift toward the Al region, which is consistent with the severe Mg/Al confusion observed in Fig.~\ref{fig:confusion_matrix}. A similar overlap is observed between U and W in the high-Z group. After adaptation with CMADA, the source and target feature distributions become better aligned, and the overlap between matched classes is reduced. Nevertheless, residual overlap remains between physically similar materials, particularly Mg/Al and U/W, suggesting that domain adaptation mitigates but does not fully resolve the intrinsic limitation imposed by similar scattering characteristics.

\section{\label{sec:Conclusion}Conclusion}
This study presents a practical method for material identification using cosmic-ray muon scattering, in which coarse momentum estimation is systematically combined with feature alignment under an unsupervised domain adaptation framework. By explicitly leveraging binned momentum information rather than relying on high-precision momentum measurements, the proposed method achieves reliable identification performance. This design choice enhances the feasibility of large-scale and long-term muon imaging applications in real-world environments.

Under the fully unsupervised target domain setting, the proposed CMADA method achieves an average recognition accuracy of 89.00\%. The results demonstrate that the proposed method enables effective material identification across materials with different atomic numbers without target domain labels. Previous studies have mainly focused on coarse identification among high-, medium-, and low-Z materials, while largely neglecting the role of momentum-related features. Further simulation results verify that the proposed domain adaptation framework can effectively mitigate the performance degradation caused by the distribution shift. This capability helps address the label bottleneck that commonly limits the practical deployment of muon imaging systems.

In future work, the assumption of uniform class distributions will be relaxed, and transfer learning methods designed for target domains with unknown and highly imbalanced material compositions will be investigated. Potential directions include class-prior estimation, adaptive reweighting, and balanced sampling schemes to further improve robustness under realistic deployment conditions. In addition, the identification of high-Z materials remains fundamentally constrained by the intrinsic saturation behaviour of muon scattering. To address this limitation, future studies will explore physics-informed feature fusion and multi-model representations, with the aim of capturing more discriminative signatures beyond scattering angle information alone. 

\section{\label{sec:ACKNOWLEDGMENTS}ACKNOWLEDGMENTS}

This work is supported by the Research Program of State Key Laboratory of Heavy Ion Science and Technology, Institute of Modern Physics, Chinese Academy of Sciences, under Grant No. HIST2025CS06, the National Natural Science Foundation of China (Grant No. 12105327, 12475106), the Guangdong Basic and Applied Basic Research Foundation, China (Grant No. 2023B1515120067), the self-initiated research fund of Advanced Energy Science and Technology, Guangdong Laboratory (DJL2025C010), and the Fundamental Research Funds for the Central Universities under Grant No. JZ2025HGTG0252. Computing resources were mainly provided by the supercomputing system in the Dongjiang Yuan Intelligent Computing Center.

\FloatBarrier

\bibliographystyle{unsrt}
\bibliography{refs}

\end{document}